\documentclass[prd,superscriptaddress,preprint]{revtex4}

\usepackage{mathrsfs,amsmath}
\usepackage{eurosym}
\usepackage{amssymb}
\usepackage{bbold}

\usepackage{latexsym,epsfig,epstopdf}
\usepackage{amssymb,amsmath,amsthm}

\usepackage{times}
\usepackage{color}

\usepackage{lmodern}
\usepackage[utf8]{inputenc}

\usepackage{graphicx}

\usepackage{color}

\usepackage{braket}

\usepackage[mathscr]{euscript} 

\def\be{\begin{equation}}
\def\ee{\end{equation}}
\def\bea{\begin{eqnarray}}
\def\eea{\end{eqnarray}}

\begin{document}

\title{Topological invariant of velocity field in quantum systems}
\author{Annan Fan}
\affiliation{School of Physics, Sun Yat-Sen University, Guangzhou, 510275, China}
\author{Shi-Dong Liang}
\altaffiliation{Email: stslsd@mail.sysu.edu.cn}
\affiliation{School of Physics, Sun Yat-Sen University, Guangzhou, 510275, China}

\date{\today }
\begin{abstract}
We introduce the velocity field of the Bloch electrons and propose the velocity field approach to characterize the topological invariants of quantum states.
We find that the zero modes of the velocity field flow play the roles of effective topological charges or defects.
A key global property of the zero modes is topological invariant against the parameter deformation. These can be characterized by the Euler characteristic based on the Poincar\'{e}-Hopf theorem. We demonstrate the validity of this approach by using the quantum sphere and torus models. The topological invariants of the velocity field in the quantum sphere and torus are consistent with the mathematical results of the vector fields in the manifolds of the sphere and torus, Euler characteristic $\chi=2$ for sphere and $\chi=0$ for torus. We also discuss the  non-Hermitian quantum torus model and compare differences in the topological invariants obtained using the velocity field and Chern number methods. The topological invariant characterized by the velocity field is homeomorphic in the Brillouin zone and the subbase manifold of the SU(2)-bundle of the system, whereas the Chern number characterizes a homotopic invariant that is associated with the exceptional points in the Brillouin zone. These results enrich the topological invariants of quantum states and provide novel insights into the topological invariants of quantum states.
\end{abstract}

\maketitle



\section{Introduction}
Topological invariants of quantum states exhibit novel phenomena and potential applications in condensed matter physics and quantum computation.\cite{Ramy,Chiu,Qi,Gong,Kohei}
A lot of effort has been focused on the classification of the topological equivalence classes based on the symmetries of different systems,\cite{Chiu,Kane1} and  on determining physical phenomena emerge in different topological phases. \cite{Qi,Yu,Annan}
It has been found that there are 10-topological equivalence classes for Hermitian systems based on the Altland-Zirnbauer(AZ)symmetry classification and 38-topological equivalence classes for non-Hermitian systems due to additional sublattice symmetry and pseudo-Hermticity. \cite{Gong,Kohei}

Non-Hermitian systems contain complex eigen energies and nonorthogonal eigenstates. This prompts many attempts to explore novel physical phenomena beyond those present in conventional Hermitian systems and novel mathematical structures beyond those of canonical quantum mechanics.\cite{Kohei,Ali}

A family of quantum Hall effects exhibit the geometric and topological properties of quantum states, and can be described using the Berry phase, winding number and Chern number.\cite{Chiu} These discoveries inspire attempts to explore fundamental issues regarding what physical features of quantum states lie behind these geometric and topological properties.\cite{Chiu,Qi}
In particular, quantum Hall conductance can be generalized to quantum Hall admittance in the non-Hermitian Dirac model.\cite{Annan} This implies that quantum Hall susceptance, such as quantum Hall capacity and induction, can emerge in non-Hermitian systems. This discovery provides a novel and fundamental insight into non-Hermitian systems and their potential applications.\cite{Annan}

One of the important features of topological phases is that the associated topological invariants of quantum states usually involve the boundary states and the energy band gap near the Fermi energy. This is called as the bulk-boundary correspondence. The non trivial topological invariants depend on the features of the boundary states near the Fermi energy in the gapped modes.\cite{Chiu,Kane1,Lee,Yao} The energy band gap based on the symmetries of the system protects the topological invariants against parameter deformation.\cite{Chiu,Ghatak,Zhang}

From a mathematical point of view, topology provides a precise tool to study the global behavior of geometric objects. Many different topological indices or numbers are used to characterize different topological invariants of geometric objects, such as the intersection number, winding number, Chern number and Euler characteristic.\cite{Eber} They label the homeomorphic equivalence classes for different geometric objects.\cite{Eber}

Many approaches have focused on the geometric properties of the wave function, such as the Berry phase and Berry curvature. They are expressed in terms of a gauge potential and gauge field,\cite{Liang,Bohm,Ghatak} which are associated with topological indices, such as the winding number and Chern number.\cite{Chiu,Ghatak,Zhang}
Recently, the wrapping number was introduced as a unified approach to obtain the topological invariants and to give the correlation length, universality classes, and scaling laws, which are related to the topological phase transition of the Dirac models in arbitrary dimensions and for arbitrary symmetry classes.\cite{Chen1}

The vorticity of the quantum state was also defined as the complex angle of the complex energy band structure as an additional way to characterize the topological invariants of non-Hermitian systems. \cite{Ghatak,Fu} In the Hamitian systems, the vorticity is equivalent to the winding number, but is defined in the complex energy space for non-Hermitian systems. \cite{Kohei,Fu} In particular, the topological defects and gapless modes in insulators and superconductors are associated with the generalized bulk-boundary correspondence.\cite{Kane2} This correspondence provides an approach to classify temporal pumping cycles, such as the Thouless charge pump, fermion parity pump and Majorana zero modes.\cite{Kane2} These results inspire us to propose a novel approach to explore topological invariants of quantum states based on the zero modes of the velocity field induced by the energy band structure.

In this paper, we propose a velocity field approach to characterize the topological invariants of quantum states. In Sec II, we introduce the velocity field of the Bloch electrons and discuss the basic properties of the velocity field in condensed matter physics. In Sec. III, we give the velocity field in the topological manifold of the quantum sphere and torus. The emergence of zero modes (or nodes) of the velocity field in the manifold play the role of effective topological charges or defects that are associated with topological invariants. We find the connection between the total indices of the zero modes and the topological invariants, which can be expressed in terms of the Euler characteristic based on the Poincar\'{e}-Hopf theorem.\cite{Eber}
In Sec. IV, we demonstrate the validity of the velocity field for our example systems. We give the topological invariants in terms of the Euler characteristic and show that results are consistent with the mathematical results of the vector field for the sphere and torus. In Sec. V, we also discuss the relationship between the topological invariants characterized by the velocity field and those characterized by the Chern number. Finally, we give a summary of our conclusions and outlook in Sec. VI. For clarity, we briefly review the Poincar\'{e}-Hopf theorem in the Appendix.

\section{The velocity field of Bloch electrons}
\subsection{ The velocity field in Hermitian systems}
Let us consider a quantum system described by a bound Hermitian Hamiltonian $H(\mathbf{k},\lambda)$ in the Brillouin zone (BZ),  where $\mathbf{k}\in BZ^{d}$ is the $d$-dimensional BZ and $\lambda\in \mathcal{M}^{p}_{\lambda}$ is a set of parameters in the $p$-dimensional parameter space.
The eigen equations are
\begin{equation}\label{HH}
H(\mathbf{k},\lambda)|\psi_{n}(\mathbf{k},\lambda)\rangle=E_{n}(\mathbf{k},\lambda)|\psi_{n}(\mathbf{k},\lambda)\rangle .
\end{equation}
Supposing that the Hilbert space of this system is separable, the eigen vectors of the Hamiltonian consist of
an orthogonal basis, $\langle\psi_{m}(\mathbf{k},\lambda)|\psi_{n}(\mathbf{k},\lambda)\rangle=\delta_{mn}$ and
the completeness relation is given by $\sum_{n}|\psi_{n}(\mathbf{k},\lambda)\left\rangle\right\langle\psi_{n}(\mathbf{k},\lambda)|=I$,
where $I$ is the identity matrix. The eigen vectors describe the wave function of the Bloch electrons and their corresponding eigen energies forms the energy bands in the BZ. In principle, most of physical properties of quantum states depend on their associated wave functions and their corresponding energy bands.
In particular, the transport properties in condensed matter physics, such as the electronic and heat conductances, depend on the velocity of the Bloch electrons. The velocity of the Bloch electron is defined  by\cite{Mermin}
\begin{equation}\label{VF0}
\mathbf{v}_{n}(\mathbf{k},\lambda)=\frac{1}{\hbar}\nabla_{\mathbf{k}}E_{n},
\end{equation}
where $\nabla_{\mathbf{k}}$ is the gradient operator for the 3D BZ.

\subsection{ The velocity field in non-Hermitian systems}
For a non-Hermitian Hamiltonian, $H^{\dag}\neq H$, the pair of eigen equations is given by
\begin{subequations}\label{HH}
\begin{eqnarray}
H(\mathbf{k},\lambda)|\psi_{n}^{R}(\mathbf{k},\lambda)\rangle=E_{n}(\mathbf{k},\lambda)|\psi_{n}^{R}(\mathbf{k},\lambda)\rangle \\
H^{\dagger}(\mathbf{k},\lambda)|\varphi_{n}^{L}(\mathbf{k},\lambda)\rangle=E^{*}_{n}(k,\lambda)|\varphi_{n}^{L}(\mathbf{k},\lambda)\rangle.
\end{eqnarray}
\end{subequations}
Similarly, let us suppose that the Hilbert space of the non-Hermitian system is separable, so that the eigen vectors of the Hamiltonian and its Hermitian adjoint form  biorthogonal basis as,\cite{Ali}
\begin{equation}\label{Orbs}
\langle\varphi_{m}^{L}(\mathbf{k},\lambda)|\psi_{n}^{R}(\mathbf{k},\lambda)\rangle=\delta_{mn},
\end{equation}
where  $|\psi_{n}^{R}(\mathbf{k},\lambda)\rangle$ and $\langle\varphi_{n}^{L}(\mathbf{k},\lambda)|$ are the corresponding eigenstates of the Hamiltonian and its Hermitian adjoint. They are labeled by $R$ and $L$, respectively. The completeness relation is given by
\begin{equation}\label{CPR1}
\sum_{n}|\psi_{n}^{R}(\mathbf{k},\lambda)\left\rangle\right\langle\varphi_{n}^{L}(\mathbf{k},\lambda)|=I,
\end{equation}
In general, the eigen energies are complex. The velocity of the Bloch electrons is defined as
\begin{equation}\label{VF2}
\mathbf{v}^{\alpha}_{n}(\mathbf{k},\lambda)=\frac{1}{\hbar}\nabla_{\mathbf{k}}E_{n}^{\alpha},
\end{equation}
where $\nabla_{\mathbf{k}}$ is the gradient operator and $\alpha=R,I$ denote the real and imaginary parts of the velocities and the energy bands, respectively.

The velocity of the Bloch electron can be expressed in terms of a covector field in the BZ, namely $\mathbf{v}^{\alpha}_{n}(\mathbf{k},\lambda)=d E_{n}^{\alpha}$, where we set $\hbar=1$ without loss of generality in the following section.
We study the analytic properties of the velocity field in the BZ to reveal novel physical properties of the system. The imaginary part of the velocity field describes the energy and probability flows of the Bloch electrons. It is found to play a crucial role in determining the physical properties of non-Hermitian systems.

\subsection{The density of states, the van Hove singularity and effective topological charges}
Let us first review the basic physical properties of the velocity field, such as the density of states and its corresponding transport properties.\cite{Mermin}

Note that the mathematical identity, $\nabla\times \nabla \mathbf{a}=0$, which holds for an arbitrary 3D vector $\mathbf{a}$, the curl of the velocity field vanishes in the BZ, $\nabla_{\mathbf{k}}\times \mathbf{v}^{\alpha}_{n}=0$. This means that the velocity field is non-curl field. On the other hand, the divergence of the velocity field is not always zero, $\nabla_{\mathbf{k}}\cdot \mathbf{v}^{\alpha}_{n}\neq 0$, which implies the existence of zero modes, such as source, sink or saddle points. These singularities play the role of effective topological charges or defects associated with the topological properties of the system.

The density of states is expressed in terms of the velocity field of the Bloch electrons,\cite{Mermin}
\begin{equation}\label{DS1}
\rho^{\alpha}_{n}(E)=\int_{S^{\alpha}_{n}}\frac{dS_{n}}{4\pi^3}\frac{\hbar}{|\mathbf{v}^{\alpha}_{n}(\mathbf{k})|},
\end{equation}
where $S^{\alpha}_n$ is the energy surface of the energy level from $E^{\alpha}_n$ to $E^{\alpha}_n+\triangle E^{\alpha}_n$. For any $\mathbf{k}\in BZ$ such that, $\mathbf{v}^{\alpha}_{n}(\mathbf{k})=0$, the density of states diverges. These characteristics are known as von Hove singularity.\cite{Mermin} These singularities not only dominate the electronic and heat transport of the system, but also play the role of effective topological charges or defects that are associated with the topological invariants.

In general, the energy bands near the Fermi energy play a crucial role in the physical properties of quantum systems.
Thus, we can investigate the behavior of the velocity field of the energy band near the Fermi energy to explore their physical properties.
In particular, the imaginary part of the velocity field is related to the dissipative properties of non-Hermitian systems. In the following sections, we will show that there exists the interplay between the real and imaginary parts of the velocity field. Thus, the imaginary part of the velocity field reveals novel phenomena in non-Hermitian systems, beyond those present in Hermitian systems, such as quantum Hall admittance.\cite{Annan}

\section{Topological invariants of the velocity field}
\subsection{The velocity field on a topological manifold}
In order to explore the relationship between the velocity field of the Bloch electrons and the topological invariants, we consider a typical two-level model with the following Hamiltonian in the 2D BZ,
\begin{equation}\label{nH2}
H(\mathbf{k},\mathbf{\lambda})=\mathbf{h}(\mathbf{k},\mathbf{\lambda})\cdot \mathbf{\sigma},
\end{equation}
where $\mathbf{\sigma}=\left(\sigma_{x},\sigma_{y},\sigma_{z}\right)$ is the vector form of the Pauli matrix and $\mathbf{\lambda}\in \mathcal{M}^{p}$ is a set of parameters. The Hamiltonian is Hermitian when $\mathbf{h}(\mathbf{k},\mathbf{\lambda})=\left(h_{x}(\mathbf{k},\mathbf{\lambda}),h_{y}(\mathbf{k},\mathbf{\lambda}),h_{z}(\mathbf{k},\mathbf{\lambda})\right)\in \mathbb{R}^3$.
From a geometric point of view, $\mathbf{h}(\mathbf{k},\mathbf{\lambda})$ plays a role of the submanifold of the base manifold of the SU(2)-bundle (mathematically, the base manifold of the SU(2)-bundle is $\mathbb{R}^{3}$),\cite{Eber}
\begin{equation}\label{TF}
\mathbf{h}(\mathbf{k},\mathbf{\lambda})
=\left\{(h_{x}(\mathbf{k},\mathbf{\lambda}),h_{y}(\mathbf{k},\mathbf{\lambda}),h_{z}(\mathbf{k},\mathbf{\lambda}))|\mathbf{k}\in(0,2\pi)^2\right\},
\end{equation}
where the BZ is two-dimensional (2D). In (\ref{TF}), $\mathbf{h}(\mathbf{k},\mathbf{\lambda})$
can actually be regarded as the 2D surface of a compact manifold embedded in $\mathbb{R}^{3}$ for a given set of  parameters $\lambda\in\mathcal{M}^{p}$.

In general, the velocity field can be redefined as a vector field \cite{Eber}
\begin{equation}\label{VF2}
\mathbf{v}(\mathbf{k},\mathbf{\lambda})  = v_x(\mathbf{k},\mathbf{\lambda})\frac{\partial \mathbf{h}}{\partial k_x}
+v_y(\mathbf{k},\mathbf{\lambda})\frac{\partial \mathbf{h}}{\partial k_y}
\end{equation}
in the manifold in (\ref{TF}). The individual components of the velocity field (\ref{VF2}) are given by
\begin{equation}
v_{n,x}(\mathbf{k},\mathbf{\lambda}) = \frac{\partial E_n}{\partial k_x},
\quad  v_{n,y}(\mathbf{k},\mathbf{\lambda})=\frac{\partial E_n}{\partial k_y}
\end{equation}
where
\begin{subequations}\label{CBMf1}
\begin{eqnarray}
\frac{\partial\mathbf{h}}{\partial k_x} &=& \left(\frac{\partial h_{x}}{\partial k_x},\frac{\partial h_{y}}{\partial k_x},\frac{\partial h_{z}}{\partial k_x}\right) \\
\frac{\partial\mathbf{h}}{\partial k_y} &=& \left(\frac{\partial h_{x}}{\partial k_y},\frac{\partial h_{y}}{\partial k_y},\frac{\partial h_{z}}{\partial k_y}\right)
\end{eqnarray}
\end{subequations}
are the basis vectors in the manifold in (\ref{TF}).

Note that the velocity field in the BZ is the local representation of the vector field in the manifold. In other words, we can infer the topological invariants of the manifold from the analytic behavior of the velocity field in the BZ when the basis (\ref{CBMf1}) in the manifold is compatible with the basis in the BZ, they are linearly independent and can be expressed as
\begin{equation}\label{CTJD1}
\frac{\partial\mathbf{h}}{\partial k_x}\times \frac{\partial\mathbf{h}}{\partial k_y}\neq 0.
\end{equation}
Thus, the topological invariants of the manifold can be characterized by the velocity field in the BZ.

\subsection{The Poincar\'{e}-Hopf theorem and topological invariants}

Due to the particle-hole symmetry of the Hamiltonian in (\ref{nH2}), the velocity field is independent of the energy band indices. Therefore, we ignore the band indices without loss of generality in the following section. When the velocity field contains finite isolated regular zero modes $\mathbf{k}_0$, such that $\mathbf{v}(\mathbf{k}_0)=0$,
the index of the zero modes is defined by,\cite{Eber}
\begin{equation}\label{Ind1}
I_{d} \left[\mathbf{v'}(\mathbf{k}_0,\lambda)\right]:=(-1)^{S_{g}} ,
\end{equation}
where $S_{g}$ depends on the determinant of the velocity field,
\begin{equation}\label{JDVF1}
\det \left[\mathbf{v'}(\mathbf{k}_0,\lambda)\right]:=
\begin{vmatrix}
\frac{\partial v_{x}(\mathbf{k},\lambda)}{\partial k_{x}} & \frac{\partial v_{x}(\mathbf{k},\lambda)}{\partial k_{y}} \\
\frac{\partial v_{y}(\mathbf{k},\lambda)}{\partial k_{x}} & \frac{\partial v_{y}(\mathbf{k},\lambda)}{\partial k_{y}}
\end{vmatrix}_{\mathbf{k}=\mathbf{k}_{0}}.
\end{equation}
$S_{g}=0$ for $\det\left[\mathbf{v'}(\mathbf{k}_0,\lambda)\right]>0$ and $S_{g}=1$ for $\det\left[\mathbf{v'}(\mathbf{k}_0,\lambda)\right]<0$.
In other words, $S_g$ is the sign of the determinant of the velocity field at the zero mode. \cite{Eber} The index of the zero mode is
$I_{d} \left[\mathbf{v'}(\mathbf{k}_0,\lambda)\right]=1$ for a sink or source, whereas $I_{d} \left[\mathbf{v'}(\mathbf{k}_0,\lambda)\right]=-1$ for a saddle point of the velocity field flows near the zero modes.\cite{Eber}

Mathematically, the Poincar\'{e}-Hopf theorem provides a description of the topological invariants of the velocity field in manifolds, which are characterized by the Euler characteristic. The Poincar\'{e}-Hopf theorem claims that the velocity field $\mathbf{v}$, as a vector field on the smooth manifold (\ref{TF}) contains finite zero modes. When the manifold is compact and orientable, the sum of indices of these zero modes is equal to the Euler characteristic,\cite{Eber}
\begin{equation}\label{PHT}
\sum_{j}I_{d} \left[\mathbf{v}(\mathbf{k}_{0,j},\lambda)\right]=\chi(\lambda),
\end{equation}
where the sum $j$ covers all the zero modes in the BZ.
This global property of the indices of the velocity field implies that the topological invariants of the manifold are characterized by the Euler characteristic. In general, the topological invariants depend on the parameters, $\lambda\in\mathcal{M}^p$, such that we can obtain the topological phase diagram by varying the parameters in the parameter space.

In general, the velocity field can characterize the topological invariants of the manifold based on the conditions of the Poincar\'{e}-Hopf theorem,\cite{Eber} which are
(1) the manifold in (\ref{TF}) is continuous to $C^\ell$ with $\ell\geq 2$, compact and orientable (non-degenerate); (2) the basis in (\ref{CBMf1}) is compatible with the basis in the BZ, i. e., they are linearly independent (satisfying (\ref{CTJD1})); (3) the zero modes are isolated and finite.
When these conditions are fulfilled for any physical system, the topological invariants of the velocity field in the BZ are equivalent to the topological invariants in the manifold of the system. These topological invariants can be characterized by the Euler characteristic based on the Poincar\'{e}-Hopf theorem. Mathematically, the covector of the velocity field in the BZ is the pull-back of the velocity field in the manifold. They are isomorphic each other.

For non-Hermitian systems, namely $\mathbf{h}(\mathbf{k},\mathbf{\lambda})\in \mathbb{C}^3$, the manifold in (\ref{TF}) becomes complex. The situation becomes complicated even though the manifold can be expressed in terms of the real and imaginary parts, respectively. The real and imaginary parts of the velocity field involve both the real and imaginary parts of $\mathbf{h}(\mathbf{k},\mathbf{\lambda})$, which can violate the above conditions. In the following section we will discuss the topological invariants of the non-Hermitian torus model and explore how to connect them to the Euler characteristic.

In principle, the velocity field approach can be generalized to higher dimensional systems since the Poincar\'{e}-Hopf theorem is applicable to the nD compact and orientable manifolds.

\begin{figure}[htbp]
	\centering
	\includegraphics[width=1.4\linewidth]{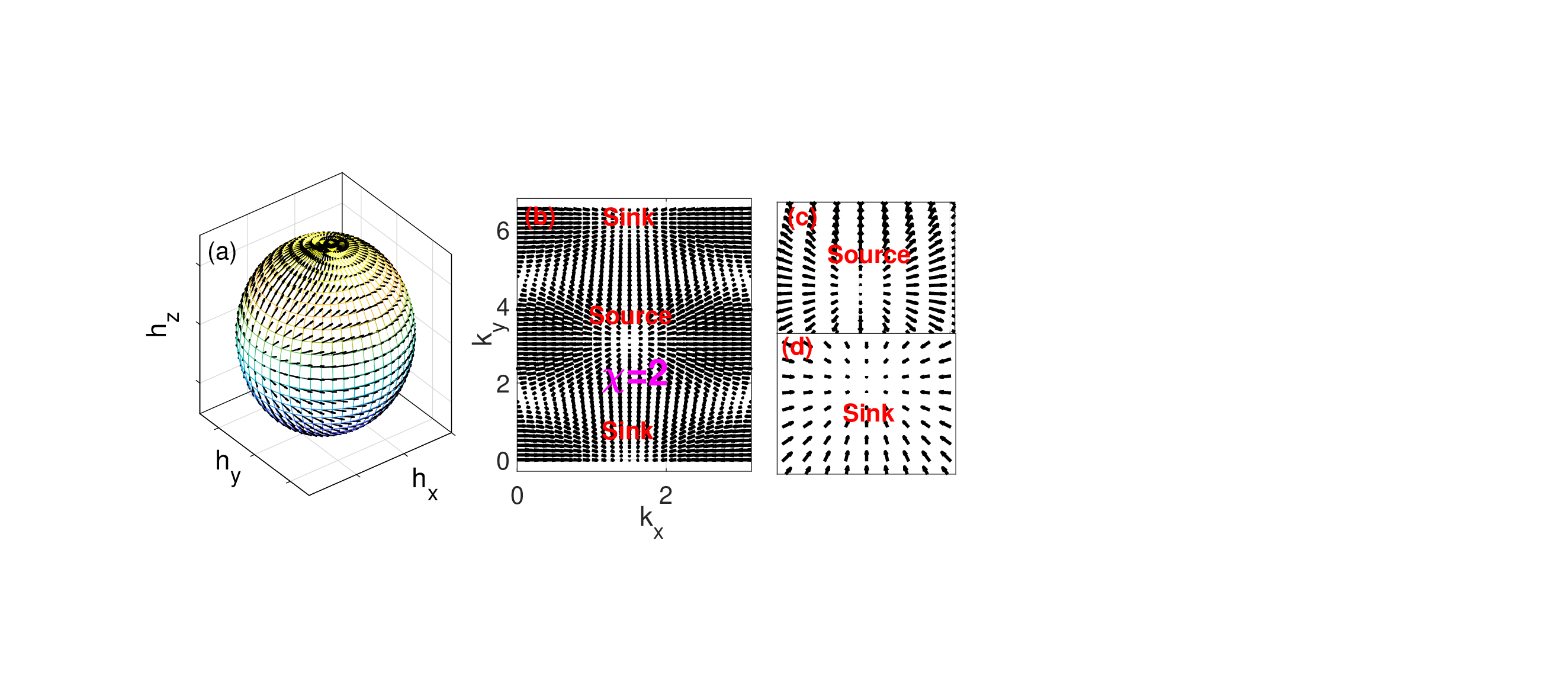}
	\caption{ Online Color: (a) The velocity field on the quantum sphere. (b) The velocity field in the BZ, where $r=5$ and a=1. (c) and (d) show the detailed flows of the source and sink respectively.}
	\label{fig1}
\end{figure}

\section{Topological invariant of quantum systems}
\subsection{The quantum sphere}
As a typical example of a two-level system, let us consider a quantum sphere model as a toy model to test the velocity field approach. The Hamiltonian is given by
\begin{equation}\label{QS1}
H(\mathbf{k},\mathbf{\lambda})=\mathbf{h}(\mathbf{k},\mathbf{\lambda})\cdot \mathbf{\sigma},
\end{equation}
where
\begin{subequations}\label{hhh1}
\begin{eqnarray}
h_{x}(\mathbf{k},\mathbf{\lambda}) &=& r\sin k_x\cos k_x+a,\\
h_{y}(\mathbf{k},\mathbf{\lambda}) &=& r\sin k_x\sin k_y\\
h_{z}(\mathbf{k},\mathbf{\lambda}) &=& r\cos k_x,
\end{eqnarray}
\end{subequations}
with $0\leq k_x\leq \pi$ and $0\leq k_y\leq 2\pi$. The parameter $r$ is the radius of the sphere and $a>0$ is a constant which shifts the $h_x$-axis, explicitly exhibiting the velocity field flow on the sphere. The 2D surface function in (\ref{hhh1}) is regarded as a submanifold of the base manifold of the SU(2)-bundle.
In other words, the submanifold is a typical sphere $S^2$. We will refer to the submanifold as simply as "manifold" for convenience, from here on.
The energy bands of this model are obtained as,
\begin{equation}\label{EQT1}
E_{\pm}=\pm\sqrt{h^{2}_x+h^{2}_y+h^{2}_z}.
\end{equation}
The velocity field in the BZ is given by
\begin{subequations}\label{vfqt1}
\begin{eqnarray}
v_{x}(\mathbf{k},\mathbf{\lambda}) &=& \widehat{\mathbf{h}}\cdot \frac{\partial \mathbf{h}}{\partial k_x},\\
v_{y}(\mathbf{k},\mathbf{\lambda}) &=& \widehat{\mathbf{h}}\cdot \frac{\partial \mathbf{h}}{\partial k_y},
\end{eqnarray}
\end{subequations}
where $\widehat{\mathbf{h}}=\frac{\mathbf{h}}{h}$ is the unit vector of $\mathbf{h}$ and $h=\sqrt{h^{2}_x+h^{2}_y+h^{2}_z}$. We set $\hbar=1$ and ignore the sign of the energy band index $\pm$ of the velocity field without loss of generality. We then have
\begin{subequations}\label{vfqt2}
\begin{eqnarray}
\frac{\partial \mathbf{h}}{\partial k_x} &=& r \cos k_x \cos k_y \mathbf{e}_x+r \cos k_x \sin k_y \mathbf{e}_y-r \sin k_x \mathbf{e}_z,\\
\frac{\partial \mathbf{h}}{\partial k_y} &=& -r \sin k_x \sin k_y \mathbf{e}_x+r \sin k_x \cos k_y \mathbf{e}_y,
\end{eqnarray}
\end{subequations}
where $\mathbf{e}_{\alpha}$ with $\alpha=x,y,z$ are the basis vectors of the manifold.
Using (\ref{vfqt1}), the velocity field in the BZ is expressed as
\begin{subequations}\label{vfqt3}
\begin{eqnarray}
v_{x}(\mathbf{k},\mathbf{\lambda}) &=& \frac{ar}{h} \cos k_x \cos k_y,\\
v_{y}(\mathbf{k},\mathbf{\lambda}) &=& -\frac{ar}{h} \sin k_x \sin k_y,
\end{eqnarray}
\end{subequations}
where $h=\sqrt{r^{2}+c^{2}+2cr\sin k_x \cos k_y}$.
The local representation of the velocity field in the manifold is given by
\begin{equation}\label{VF3}
\mathbf{v}(\mathbf{k},\mathbf{\lambda})  =  v_x(\mathbf{k},\mathbf{\lambda})\frac{\partial \mathbf{h}}{\partial k_x}
+v_y(\mathbf{k},\mathbf{\lambda})\frac{\partial \mathbf{h}}{\partial k_y}.
\end{equation}
Using (\ref{vfqt2}), we obtain
\begin{equation}\label{CTJD2}
\frac{\partial\mathbf{h}}{\partial k_x}\times \frac{\partial\mathbf{h}}{\partial k_y}=
r^{2}\sin k_x (\sin k_x \cos k_y \mathbf{e}_x+\sin k_x \sin k_y \mathbf{e}_y+\cos k_x \mathbf{e}_z).
\end{equation}
It can be seen that $\frac{\partial\mathbf{h}}{\partial k_x}\times \frac{\partial\mathbf{h}}{\partial k_y}\neq 0$ except for the south and north poles of the sphere, $k_x=0,\pm \pi$. This implies that the basis transformation between the BZ and the manifold is compatible except for the south and north poles of the sphere.
Thus, we can ignore the zero modes at the south and north poles, namely $k_x=0,\pi$, such that the Poincar\'{e}-Hopf theorem still applies.

Note that the manifold is a typical sphere $S^{2}$, which is compact and orientable. The velocity field is continuous except for the south and north poles of the sphere. We plot the manifold of the quantum sphere and the velocity field in the BZ in Fig. \ref{fig1}, in which we neglect the zero modes in the BZ boundary $k_x=0,\pi$ (the south and north poles). We can see that one source is located at the point $(\pi/2,\pi)$ and two sinks are located at the points $(0,2\pi)$ and $(\pi/2,2\pi)$, respectively. The total index of the zero modes is obtained as
\begin{eqnarray}\label{IndS}
I_{d} \left[\mathbf{v}^{R}(\mathbf{k}_0,\lambda)\right] &=& (-1)^{S_{source}}+2\frac{1}{2}(-1)^{S_{sink}}=1+1=2=\chi,
\end{eqnarray}
where the last equality follows from the Poincar\'{e}-Hopf theorem.
This gives the indices of the zero modes of the velocity field and the Euler characteristic $\chi=2$. This result is consistent with the known topology of a larger class than spheres, $\chi=2-2g=2$, where $g=0$ for spheres. The quantum sphere here is regarded as a toy model to test the validity of the velocity field approach.

It should be remarked that the topological invariants of the velocity field are independent of the parameter $a$ even though the local flow depends on the parameter $a>0$. The different observables of the local and global descriptions of the velocity field will be studied in further works.

On the other hand, the Berry curvature is given by,\cite{Chiu, Annan}
\begin{equation}\label{BC1}
\Omega= \frac{1}{2}\mathbf{\widehat{h}}\cdot
\left(\frac{\partial \mathbf{\widehat{h}}}{\partial k_x}\times\frac{\partial \mathbf{\widehat{h}}}{\partial k_y}\right).
\end{equation}
Using (\ref{vfqt2}), the Berry curvature is obtained as
\begin{equation}\label{BC2}
\Omega(\mathbf{k},\lambda)= \frac{r^2}{2h^3}(r+a \sin k_x \cos k_y )\sin k_x .
\end{equation}
Thus, the Chern number of the quantum sphere is given by
\begin{equation}\label{CNqt1}
C (\lambda)= \frac{1}{2\pi}\int_{\textrm{BZ}}\Omega(\mathbf{k},\lambda)d^{2}\mathbf{k}=1,
\end{equation}
which is independent of $a$ except for the gapless modes, $h=0$. In the gapless modes, the Chern number becomes ill-defined. Comparing the topological invariants obtained by the velocity field and those obtained from the Chern number provides a way to test the validity of the velocity field approach.

\begin{figure}[htbp]
	\centering
	\includegraphics[width=1.4\linewidth]{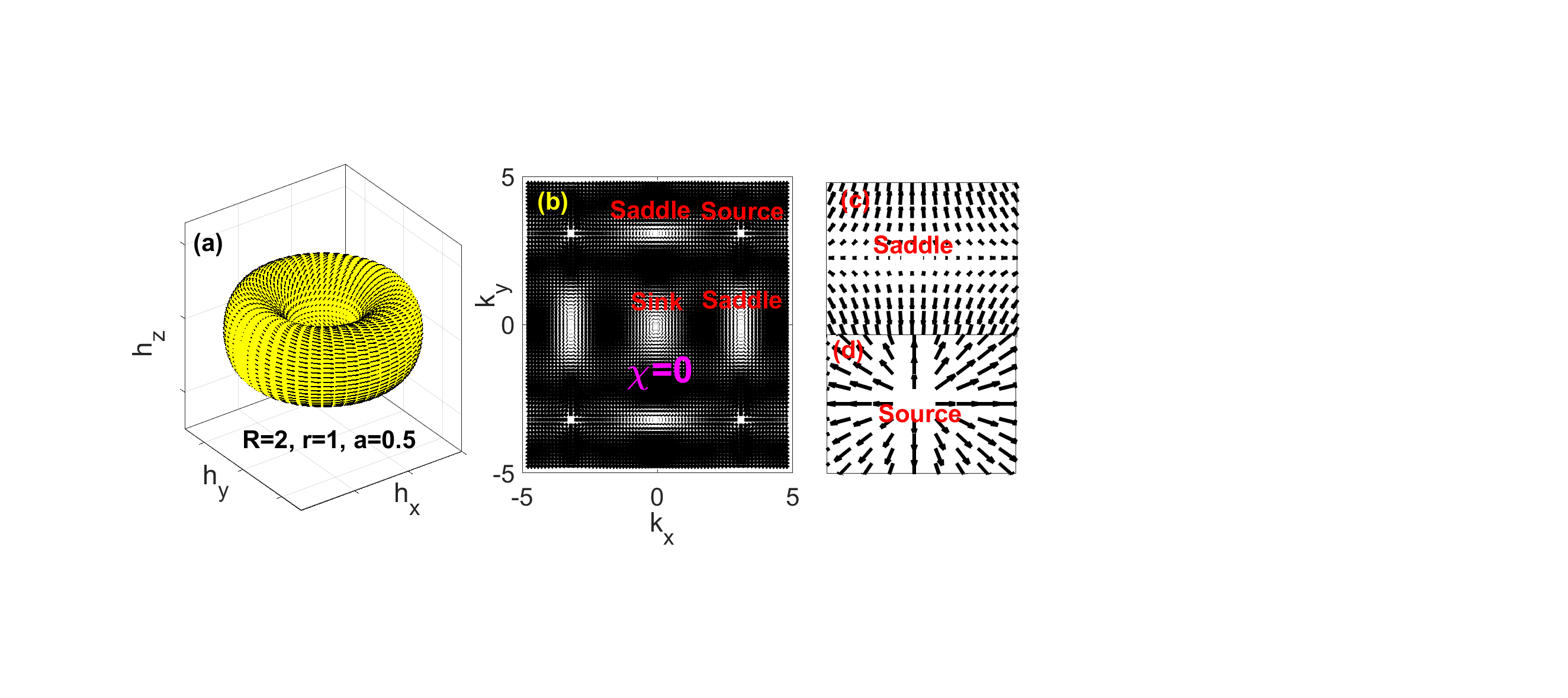}
	\caption{ Online Color: (a) The velocity field on the quantum torus. (b) The velocity field in the BZ, where $R=2$, $r=1$ and a=1.(c) and (d) show the detailed flows of the source and sink respectively.}
	\label{fig2}
\end{figure}

\subsection{The quantum torus}
Another typical example of two level systems is the quantum torus model. The Hamiltonian is given by
\begin{equation}\label{QT3}
H(\mathbf{k},\mathbf{\lambda})=\mathbf{h}(\mathbf{k},\mathbf{\lambda})\cdot \mathbf{\sigma},
\end{equation}
where
\begin{subequations}\label{hhh3}
\begin{eqnarray}
h_{x}(\mathbf{k},\mathbf{\lambda}) &=& r_{0}\cos k_{x}+a,\\
h_{y}(\mathbf{k},\mathbf{\lambda}) &=& r_{0}\sin k_{x},\\
h_{z}(\mathbf{k},\mathbf{\lambda}) &=& r\sin k_{y},
\end{eqnarray}
\end{subequations}
with $r_0\equiv\sqrt{r^{2}\sin^{2}k_y+(R+r\cos k_y)^2}$. The parameter $R$ is the radius of the torus and $r$ is the radius of the ring, $R>r$. The constant $a$ shifts the $h_x$-axis of the torus ($0<a<r$),  explicitly exhibiting the velocity field flow. The wave vector is within the 2D BZ, $-\pi\leq k_x,k_y\leq \pi$.
Similarly, the 2D surface function in (\ref{hhh3}) is regarded as a submanifold (of the base manifold of the SU(2)-bundle). This implies that the submanifold is a typical torus $T^2$. The energy bands of the model are
\begin{equation}\label{EQT1}
E_{\pm}=\pm\sqrt{h^{2}_x+h^{2}_y+h^{2}_z}.
\end{equation}
As with the quantum sphere, the velocity field in the BZ is given by
\begin{subequations}\label{vf3}
\begin{eqnarray}
v_{x}(\mathbf{k},\mathbf{\lambda}) &=& \widehat{\mathbf{h}}\cdot \frac{\partial \mathbf{h}}{\partial k_x},\\
v_{y}(\mathbf{k},\mathbf{\lambda}) &=& \widehat{\mathbf{h}}\cdot \frac{\partial \mathbf{h}}{\partial k_y},
\end{eqnarray}
\end{subequations}
then we have
\begin{subequations}\label{vfqt3}
\begin{eqnarray}
\frac{\partial \mathbf{h}}{\partial k_x} &=& -r_0 \sin k_x \mathbf{e}_x+r_0 \cos k_x  \mathbf{e}_y,\\
\frac{\partial \mathbf{h}}{\partial k_y} &=& -\frac{rR}{r_0} \sin k_y (\cos k_x \mathbf{e}_x+ \sin k_x \mathbf{e}_y)+r \cos k_y \mathbf{e}_z.
\end{eqnarray}
\end{subequations}
Using (\ref{vf3}), the velocity field in the BZ can be obtained as
\begin{subequations}\label{VFxy2}
\begin{eqnarray}
v_{x}(\mathbf{k},\mathbf{\lambda}) &=& -\frac{r_{0}c\sin k_x}{h},\\
v_{y}(\mathbf{k},\mathbf{\lambda}) &=& -\frac{rR}{h}\left(1+\frac{a}{r_0}\cos k_{x}-\frac{r}{R}\cos k_y\right)\sin k_y.
\end{eqnarray}
\end{subequations}
The zero modes are given by the conditions $v_{x}(\mathbf{k},\mathbf{\lambda}) =v_{y}(\mathbf{k},\mathbf{\lambda}) =0$. It can be found from (\ref{VFxy2}) that the zero modes are located at $k_x=0,\pm \pi$ and $k_y=0,\pm \pi$.

The local representation of the velocity field in the manifold of the quantum torus is
\begin{equation}\label{VF4}
\mathbf{v}(\mathbf{k},\mathbf{\lambda})= v_x(\mathbf{k},\mathbf{\lambda})\frac{\partial \mathbf{h}}{\partial k_x}
+v_y(\mathbf{k},\mathbf{\lambda})\frac{\partial \mathbf{h}}{\partial k_y}.
\end{equation}
It is easy to check that $\frac{\partial \mathbf{h}}{\partial k_x}\cdot\frac{\partial \mathbf{h}}{\partial k_y}=0$ in the BZ. This implies that the basis in the manifold is orthogonal and the transformation between the BZ and the manifold, $\left(\frac{\partial \mathbf{k}}{\partial k_{x}},\frac{\partial \mathbf{k}}{\partial k_{y}}\right)\rightarrow\left(\frac{\partial \mathbf{h}}{\partial k_{x}},\frac{\partial \mathbf{h}}{\partial k_{y}}\right)$ shows that they are compatible with each other.
Note that the manifold is a typical torus $T^{2}$ and is compact and orientable. Thus, the topological invariants of the velocity field in the torus can be characterized by the Euler characteristic based on the Poincar\'{e}-Hopf theorem.

We investigate the features of the velocity field and give their connections to the topological invariants based on the Poincar\'{e}-Hopf theorem.\cite{Eber}
We numerically plot the velocity field flow in the BZ and determine the zero modes and their indices. The global properties of the indices are described by the Euler characteristic of the manifold.

It should be noted that when a zero mode of the velocity field is located at the boundary of the BZ, the index of the velocity field is divided by two because it shares two BZs. Similarly the zero mode is located at a corner of the BZ, the index is divided by four because it shares four BZs. Due to the particle-hole symmetry of the quantum torus model, we investigate only the velocity field of the lower energy band without loss of generality.

Fig.\ref{fig2} shows (a) the velocity field on the torus and (b) is the velocity field in the BZ. We can see that one sink is located at the $\Gamma$ point $(0,0)$, source is located at each corner $(\pm \pi, \pm\pi)$, and one saddle point is located at each boundary of the BZ, $(0,\pm\pi)$ and $(\pm\pi,0)$. The total index is obtained as
\begin{eqnarray}\label{Ind1}
I_{d} \left[\mathbf{v}^{R}(\mathbf{k}_0,\lambda)\right]  &=& (-1)^{S_{sink}}+4\frac{1}{4}(-1)^{S_{source}}+4\frac{1}{2}(-1)^{S_{saddle}} \nonumber\\
   &=& 1+1-2=0=\chi,
\end{eqnarray}
which is consistent with the known topology of torus based on the Poincar\'{e}-Hopf theorem, $\chi=2-2g$,\cite{Eber} where $g$ is the genus of the manifold and $g=1$ for tori, giving $\chi=0$.

\begin{figure}[htbp]
	\centering
	\includegraphics[width=1.1\linewidth]{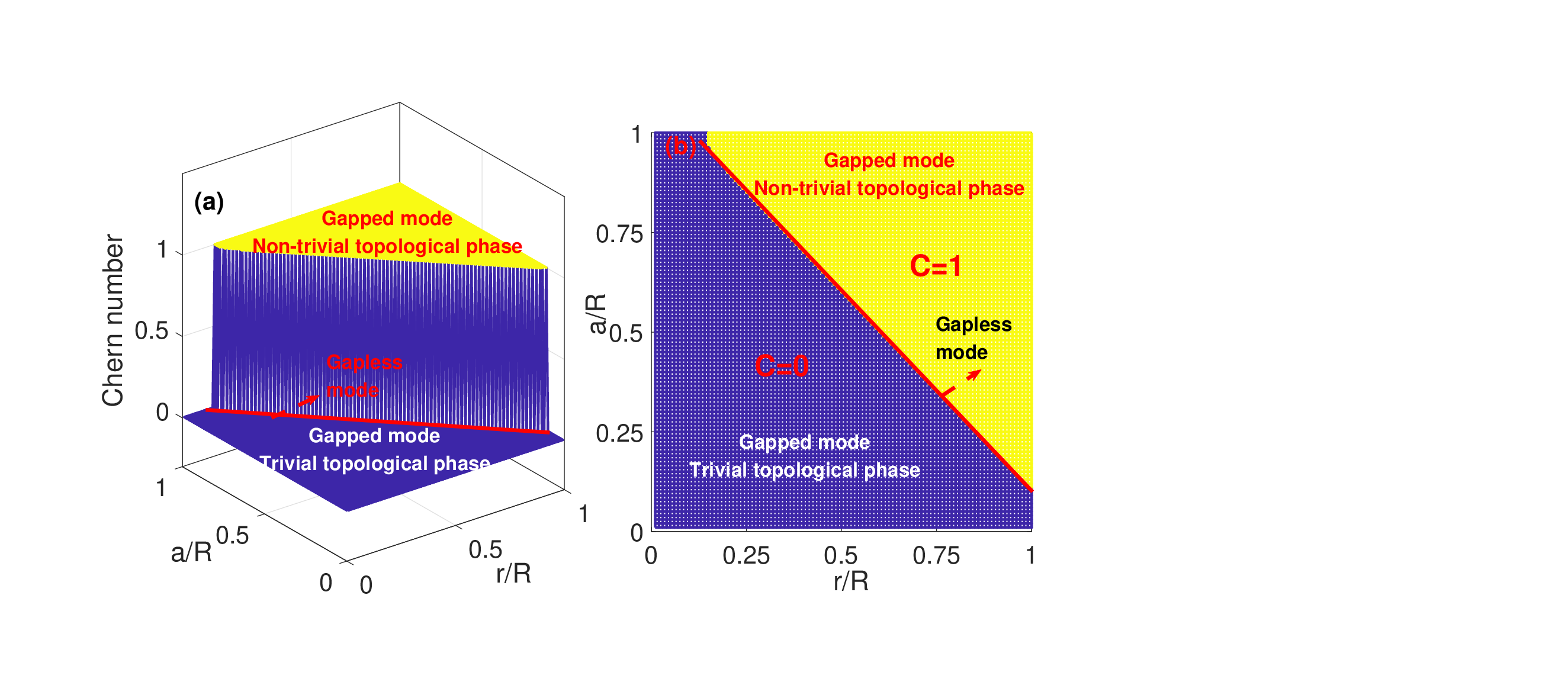}
	\caption{ Online Color: The Chern number $C$ of the quantum torus in the parameter space. The yellow region is $C=1$ for the non-trivial topological phase and the blue region is $C=0$ for the trivial topological phase of the gapped modes. The red line shows the gapless modes in the parameter space.}
	\label{fig3}
\end{figure}

Similarly, in order to compare the topological invariants of the torus based on the velocity field and the Chern number, we use (\ref{BC1}), and the Berry curvature is given by,
\begin{equation}\label{BC4}
\Omega(\mathbf{k},\lambda)= \frac{r}{2h^3}(R^2+r^2+r_0 a \cos k_x )\cos k_y +Rr^2 (\cos^{2}k_y+1).
\end{equation}
The Chern number then is obtained as
\begin{equation}\label{CNqt1}
C (\lambda)= \frac{1}{2\pi}\int_{\textrm{BZ}}\Omega(\mathbf{k},\lambda)d^{2}\mathbf{k}
\end{equation}
where $\lambda$ denotes the parameters of the torus, $R,r,a$.

Fig.\ref{fig3} shows the numerical value of the Chern number of the quantum torus based on (\ref{CNqt1}). We can see that the Chern number depends on the parameters $r$ and $a$ for a given $R$ in Fig.\ref{fig3}(a). The Chern number is either $1$ or $0$ in the $(r,a)$-parameter space. Fig.\ref{fig3}(b) shows the topological phase diagram of the torus, in which we can see that $C=1$ characterizes the non-trivial topological phase of the gapped mode and $C=0$ represents the trivial topological phase. The phase transition line (red line) between the trivial and non-trivial topological phases is the gapless mode in the parameter space. These results, based on the Chern number, are different from those based on the velocity field approach.

In principle, the velocity field characterizes the topological invariants of the manifold of the quantum torus. This topology is characterized by the Euler characteristic. The topological invariants depend only on the manifold even though the local properties of the velocity field depend on the parameters of the torus, such as the locations of the zero modes and the local flows near the zero modes. The Chern number is well-defined only for the gapped modes. The Chern number characterizes the topological phases $C=1$ for the non-trivial phase and $C=0$ for the trivial phase in the gapped modes.

\subsection{The non-Hermitian quantum torus }

Let us generalize the Hermitian quantum torus to non-Hermitian quantum torus in order to explore the topological invariants based on the velocity field approach. The Hamiltonian of the non-Hermitian quantum torus is given by
\begin{equation}\label{CS1}
H(\mathbf{k},\mathbf{\lambda})=\mathbf{h}(\mathbf{k},\mathbf{\lambda})\cdot \mathbf{\sigma}
\end{equation}
where
$\mathbf{h}(\mathbf{k},\mathbf{\lambda}) =\mathbf{h}^{R}(\mathbf{k},\mathbf{\lambda})+i\mathbf{h}^{I}(\mathbf{k},\mathbf{\lambda})\in \mathbb{C}^3$.
The real part is given by
\begin{subequations}\label{hhh4}
\begin{eqnarray}
h_{x}^{R}(\mathbf{k},\mathbf{\lambda}) &=& r_{0}\cos k_{x}+c,\\
h_{y}^{R}(\mathbf{k},\mathbf{\lambda}) &=& r_{0}\sin k_{x},\\
h_{z}^{R}(\mathbf{k},\mathbf{\lambda}) &=& r\sin k_{y},
\end{eqnarray}
\end{subequations}
and the imaginary part is
\begin{subequations}\label{hhh5}
\begin{eqnarray}
h_{x}^{I}(\mathbf{k},\mathbf{\lambda}) &=& \delta_x r_{0}\cos k_{x}+c,\\
h_{y}^{I}(\mathbf{k},\mathbf{\lambda}) &=& \delta_y r_{0}\sin k_{x},\\
h_{z}^{I}(\mathbf{k},\mathbf{\lambda}) &=& \delta_z r\sin k_{y},
\end{eqnarray}
\end{subequations}
where $r_0\equiv\sqrt{r^{2}\sin^{2}k_y+(R+r\cos k_y)^2}$.
The parameter $R$ is the radius of the torus and $r$ is the radius of the ring, $R>r$. The constant parameter $c$ shifts the $h^{R(I)}_x$-axis and we take $0<c<r$ for convenience explicitly exhibiting the velocity field flow on the torus. The parameters $\delta_{x(y,z)}$ label the determine deviations from the Hermitian quantum torus. The wave vectors are within the 2D BZ, $-\pi\leq k_x,k_y\leq \pi$, and the energy bands have the same form as is the Hermitian torus model,
\begin{equation}\label{EQT1}
E_{\pm}=\pm\sqrt{h^{2}_x+h^{2}_y+h^{2}_z}
\end{equation}
except for $E_{\pm}\in \mathbb{C}$. Similarly to the Hermitian case, the velocity field in the BZ is extended to the complex domain
\begin{subequations}\label{vf4}
\begin{eqnarray}
v_{x}(\mathbf{k},\mathbf{\lambda}) &=& \widehat{\mathbf{h}}\cdot \frac{\partial \mathbf{h}}{\partial k_x}
\equiv v_{x}^{R}(\mathbf{k},\mathbf{\lambda})+iv_{x}^{I}(\mathbf{k},\mathbf{\lambda}),\\
v_{y}(\mathbf{k},\mathbf{\lambda}) &=& \widehat{\mathbf{h}}\cdot \frac{\partial \mathbf{h}}{\partial k_y}
\equiv v_{y}^{R}(\mathbf{k},\mathbf{\lambda})+iv_{y}^{I}(\mathbf{k},\mathbf{\lambda}).
\end{eqnarray}
\end{subequations}
It should be noted that the manifold is extended to the complex domain for the non-Hermitian quantum torus. The velocity field is also generalized to become a complex vector field. In general, each of the real and imaginary parts of the complex velocity field depend on both the real and imaginary parts of the energy bands.
Using (\ref{hhh4}) and (\ref{hhh5}), we have
\begin{subequations}\label{vf5}
\begin{eqnarray}
\frac{\partial \mathbf{h}}{\partial k_x} &=& -(1+i\delta_x )r_0 \sin k_x \mathbf{e}_x+(1+i\delta_y )r_0 \cos k_x  \mathbf{e}_y,\\
\frac{\partial \mathbf{h}}{\partial k_y} &=& -\frac{rR}{r_0} \sin k_y ((1+i\delta_x )\cos k_x \mathbf{e}_x+ (1+i\delta_y )\sin k_x \mathbf{e}_y)
+(1+i\delta_z )r \cos k_y \mathbf{e}_z.
\end{eqnarray}
\end{subequations}
The complex velocity field in the BZ is then obtained as
\begin{subequations}\label{VF6}
\begin{eqnarray}
v_{x}(\mathbf{k},\mathbf{\lambda}) &=& -\frac{1}{h}\left[(1+i\delta_x )^2r^{2}_{0} \sin k_x\cos k_x-(1+i\delta_x )(1+i)cr_0\sin k_x \right. \nonumber\\
&+&\left.(1+i\delta_y )^2 r^{2}_{0}\sin k_x \cos k_x\right],\\
v_{y}(\mathbf{k},\mathbf{\lambda}) &=& -\frac{rR}{h}\sin k_y\left[(1+i\delta_x )^2  \cos^{2}k_x+(1+i)c(1+i\delta_x )\cos k_x \right. \nonumber\\
&+& \left.(1+i\delta_y )^2 r_0 \sin^{2}k_x\right]+\frac{1}{h}(1+i\delta_z)^{2}r^{2}\sin k_y \cos k_y .
\end{eqnarray}
\end{subequations}
In practice, the complex velocity field and its zero modes can be calculated by numerical methods using $v_{x}(\mathbf{k},\mathbf{\lambda}) =v_{y}(\mathbf{k},\mathbf{\lambda}) =0$. For a given set of parameters,
we can check that $\frac{\partial \mathbf{h}}{\partial k_x}\times \frac{\partial \mathbf{h}}{\partial k_y}\neq 0$, which ensures that the basis in the manifold is compatible with the basis in the BZ. The real and imaginary parts of the manifold form a typical torus $T^{2}$ and they are compact and orientable. Thus, the topological invariants of the velocity field in the manifold can be characterized by the Euler characteristic.

\begin{figure}[htbp]
	\centering
	\includegraphics[width=1.2\linewidth]{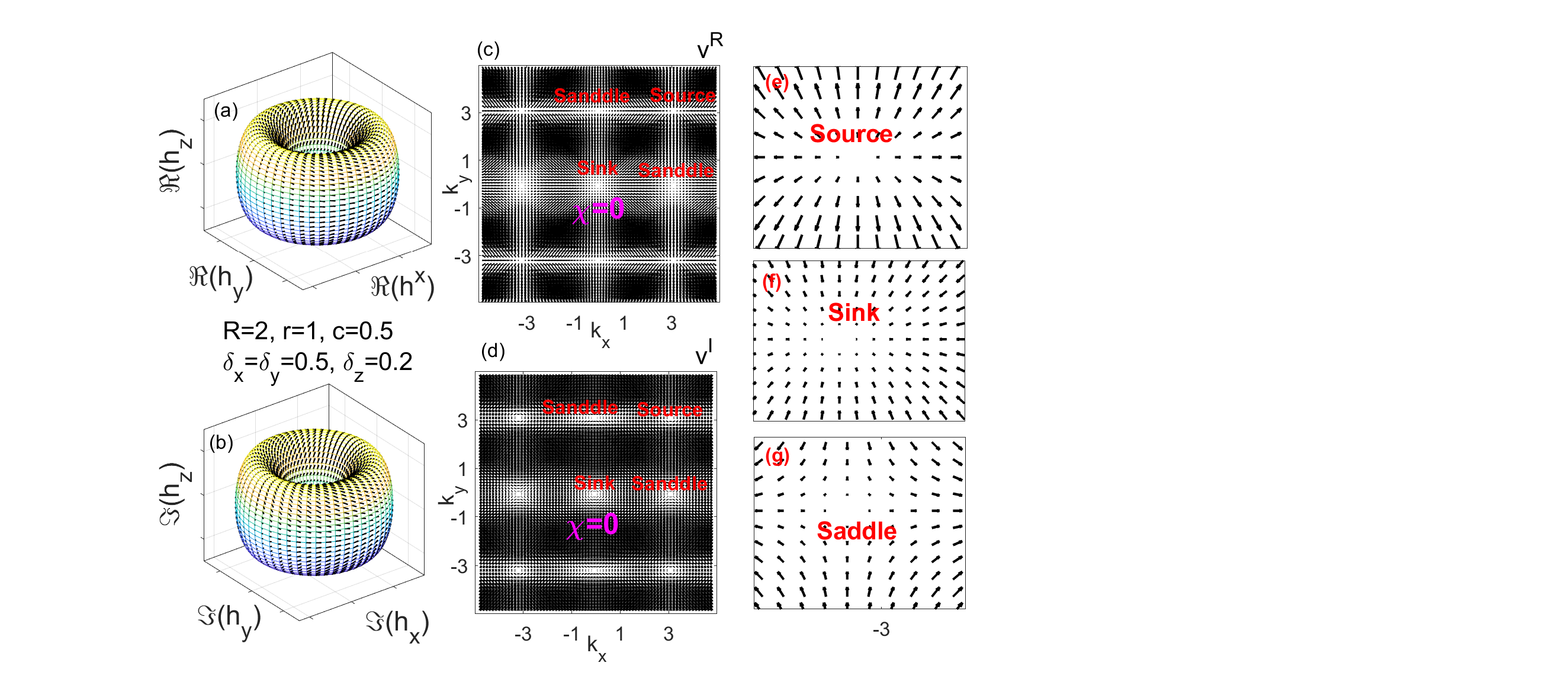}
	\caption{Online color: The non-Hermitian quantum torus and its velocity field in the BZ for the parameters, $R=2,r=1$ and $c=0.5$.
(a) and (b) show the real and imaginary parts of the torus. (c) and (d) show the real and imaginary parts of the velocity fields. The source, sink and saddle points can be seen by amplifying the figures. (e)-(f) show the detailed flows of the source, sink  and saddle points respectively.}
	\label{fig4}
\end{figure}

Let us now investigate the features of the velocity field and explore its connection to the topological invariants based on the Poincar\'{e}-Hopf theorem.\cite{Eber}
Fig.\ref{fig4} shows the non-Hermitian quantum torus and its the velocity field for the parameters $R=2,r=1$ and $c=0.5$. Fig.\ref{fig4}(a) and (b) are the real and imaginary parts of the torus, respectively, and (c) and (d) are their velocity fields in the BZ.

We can see from Fig. \ref{fig4}(c) that different types of the zero modes of the velocity field exist, including sources, sinks and saddles located symmetrically in the BZ.
One sink appears at the $\Gamma$ point $(0,0)$ inside the BZ.
Four sources are located at the M points $(\pm\pi,\pm\pi)$. Four saddles appear at the boundary points $(0,\pm\pi)$ and $(\pm\pi,0)$.
Therefore, the total index of the zero modes is expressed as
\begin{eqnarray}\label{Ind2}
I_{d} \left[\mathbf{v}^{R}(\mathbf{k}_0,\lambda)\right] &=& (-1)^{S_{sink}}+4\frac{1}{4}(-1)^{S_{source}}+4\frac{1}{2}(-1)^{S_{saddle}} \nonumber\\
   &=& 1+1-2=0=\chi,
\end{eqnarray}
Therefore, we see that our results are consistent with the Euler characteristic for tori $\chi=0$.

We now compare the topological invariants based on the velocity field and the Chern number. The Chern number of a non-Hermitian system is generalized to the complex domain as, \cite{Annan}
\begin{equation}\label{CCN1}
C_{\pm}= \frac{1}{4\pi}\int_{\textrm{BZ}}\mathbf{\widehat{h}}\cdot
\left(\frac{\partial \mathbf{\widehat{h}}}{\partial k_x}\times\frac{\partial \mathbf{\widehat{h}}}{\partial k_y}\right)d^{2}\mathbf{k},
\end{equation}

\begin{figure}[htbp]
	\centering
	\includegraphics[width=1.2\linewidth]{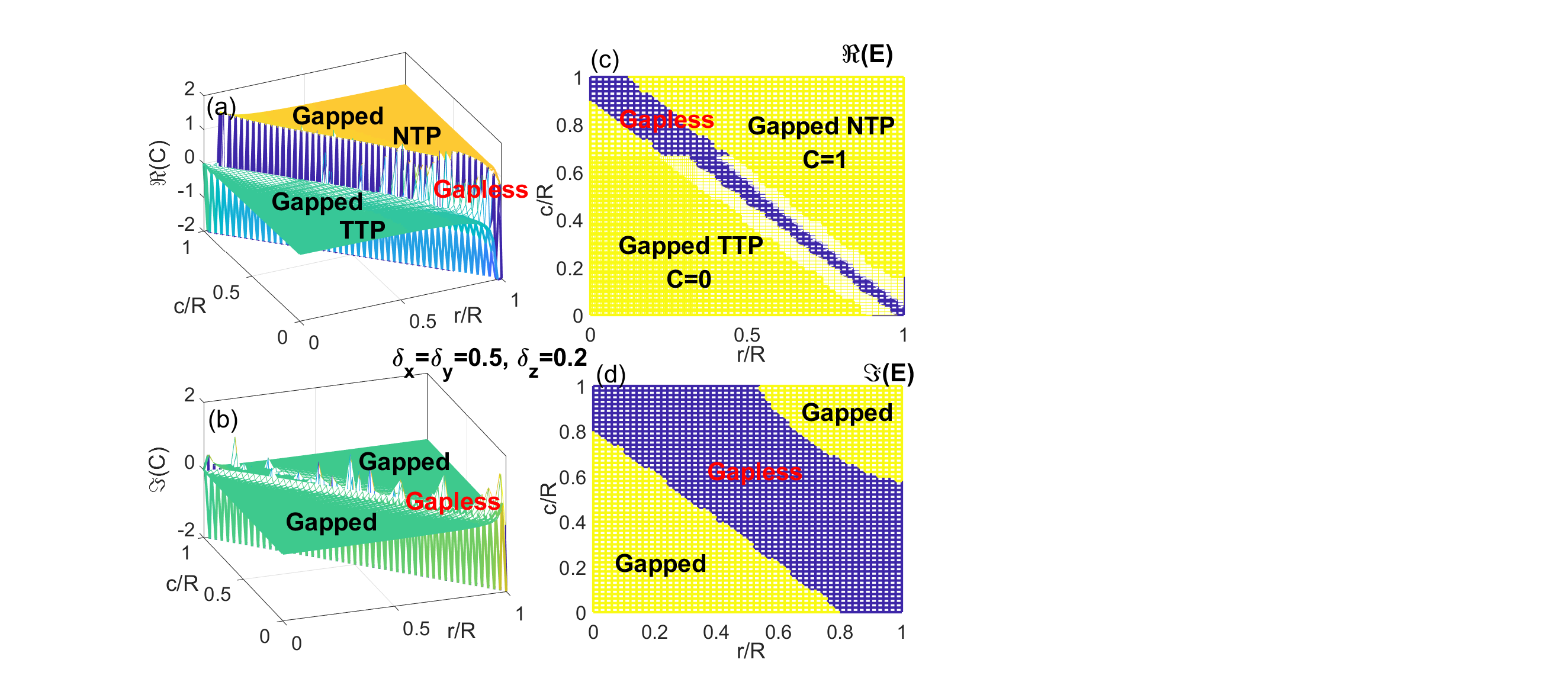}
	\caption{Online color: The Chern number of the non-Hermitian quantum torus in the parameter space for the non-Hermitian parameters $\delta_x=\delta_y=0.5, \delta_z=0.2$. (a) and (b) show the real and imaginary parts. (c) and (d) show the phase diagram in the parameter space. The acronyms NTP and TTP stand for the Non-trivial Topological Phase and Trivial Topological Phase, respectively.}
	\label{fig5}
\end{figure}

Fig.\ref{fig5}(a) and (b) shows the Chern numbers of the real and imaginary parts of the non-Hermitian quantum torus in the parameter space. We can see that the real part of the Chern number shows a step from $0$ to $1$ in Fig. \ref{fig5}(a). The Chern numbers, $C=1$ and $C=0$, characterize the non-trivial and trivial topological phases in the gapped modes, respectively. The imaginary part of the Chern number in the gapped modes is zero, but both the real and imaginary parts of the Chern number show fluctuations in the gapless modes, which can be seen in Figs. \ref{fig5}(a) and (b) because the Chern number is actually ill-defined for the gapless modes. The real and imaginary parts of the Chern number are projected in the parameter space as the phase diagram shown in Fig. \ref{fig5}(c) and (d). The region of the gapless modes for the imaginary part is larger than that of the real part.

\section{Remarks on topological invariants}
Let us compare the topological invariants obtained using the Chern number and velocity field approaches. The Chern number depends on the exceptional points of the energy band inside or outside of the 2D surface of the manifold. These correspond to the non-trivial and trivial topological invariants, respectively.
The topological invariants based on the Chern number can be regarded as homotopic function transformations within the gapped phase. The existence of exceptional points implies the closure of the energy band (gapless modes). When exceptional points emerge in the BZ, due to the variation of parameters, the Chern number becomes ill-defined.
This corresponds to the gapless modes, which can be characterized by the vorticity. \cite{Ghatak,Fu}

The topological invariants obtained based on the velocity field approach depend on the number of zero modes and the velocity field flow near the zero modes. The existence of  zero modes implies that the velocity of the Bloch electrons is zero at some points in the BZ or in the manifold. These zero modes play the role of effective topological charges or defects that dominate the topological invariants of quantum states. A global property of the indices is the Euler characteristic, based on the Poincar\'{e}-Hopf theorem. This gives a homeomorphic invariant of the manifolds of the quantum sphere and torus.

It should be noted that the concepys of a "zero mode" and an "exceptional point" are different in mathematics and physics. The topological invariants of the quantum sphere and torus characterized by the Chern number and velocity field approaches show different parameter-dependence, (on $a$ or $c$), for both of the Hermitian and non-Hermitian cases. The Chern number depends on $a$ or $c$ because the energy gap is related to either $a$ or $c$. When the energy band closes (exceptional point emerges) for a given set of the parameters $a$ or $c$, the Chern number becomes ill-defined.  This is why the topological invariants obtained by the velocity field and the Chern number show different-parameter dependences in the phase diagram.

For the velocity field approach, the existence of the Euler characteristic depends on two conditions. One is that the BZ and the manifold are
compatible such that the velocity field in the manifold for a given set of parameters $\mathbf{\lambda}$ is topologically equivalent to the velocity field in the BZ.
The other is the condition of the Poincar\'{e}-Hopf theorem, which gives the topological invariants as the Euler characteristic.
From a mathematical point of view, the velocity field in the BZ viewed as a covector field is the pull-back of the vector field in the manifold. This is why the topological invariants in the BZ are homeomorphism with those in the manifold.

For non-Hermitian cases, such as the non-Hermitian quantum torus, when the above two conditions are satisfied, the velocity field approach is still applicable as is the example case shown in Fig.(\ref{fig4}). However, when the non-Hermitian parameters are varied, e.g. such that $\delta_x \neq \delta_y$ or $c$ is moved to the $h^{I}_y$ term, the required conditions could be violated so that the topological invariants of the velocity field in the BZ are no longer equivalent to the topological invariants in the manifold and the Poincar\'{e}-Hopf theorem is also no longer applicable.
What observable physical effect occurs in this case is a worth studying in later work.

Interestingly, the velocity field describes two kinds of physical properties. One is the topological properties characterized by the total zero modes of the velocity field and the Euler characteristic. This topological invariant does not depend on the number of zero modes, their locations or the detailed velocity field flows. Namely, different numbers, locations and velocity field flows of the zero modes (source, sink, and saddle) can yield the same topological invariants.
The other is a local property of the quantum states. This local property depends explicitly on the number, locations and analytic behaviors of velocity field flows.
What observable physics is hidden behind these global and local properties is also worth studying further.

Mathematically, the Euler characteristic describes homotopically equivalent topological spaces.\cite{Eber} There is a connection between the Gaussian curvature, the Chern number and the Euler characteristic based on the Gauss-Bonnet theorem.\cite{Eber} In principle, this method can be generalized to study higher dimension and more generic models.

\section{Conclusions and outlook}
In summary, we developed a novel approach to characterize the topological invariants of the quantum states based on
the velocity field of the Bloch electrons. We found that the zero modes of the velocity field play the roles of effective topological charges or defects, which dominate the topological invariants given by the Euler characteristic based on the Poincar\'{e}-Hopf theorem.

We demonstrated the validity of the velocity field approach using the quantum sphere and torus models as examples. We plotted the numerical solutions of the velocity field and determine its zero modes, which are associated with the Euler characteristic. Our results are consistent with the known topologies of the sphere and torus. We compared and discussed the topological invariants characterized by the velocity field approach and the Chern Number based on their different physical and mathematical structures.
We found that they show different-parameter dependence because they depend on the zero modes and exceptional points, respectively.
In other words, we revealed a novel topological invariant based on the velocity field approach beyond those obtained from the Chern number.

Interestingly, the zero modes of the velocity field reveal both of local and global properties of quantum states. The local properties depend on the number and locations of the zero modes as well as their detailed flows, but the global topological invariants depends only on the total number and global properties of the zero modes. This
implies that different local behaviors of the velocity field may yield the same global topological invariants. This can be regarded as an
analog of electrodynamics. Different charge distributions inside a closed surface generate different electric fields on the surface, but the total fluxes through surface may be the same. Thus, the zero modes of the velocity field can be regarded as effective topological charges or defects.
This provides a novel way to classify the topological invariants of quantum states.
In general, the density of states depends on the velocity flow. The emergence of zero modes implies that the density of states diverges, which dominates many physical properties, such as electronic and heat transport.\cite{Mermin}

The velocity field approach provides a novel insight into the topological invariants of quantum states and which may have potential applications, but also raises many fundamental issues. What emerging physical phenomena are associated with different topological invariants characterized by the velocity field and Chern number? Especially what is the relationship between the local velocity field flow and the global topological invariant?
Moreover, the validity of the velocity field approach depends on two conditions. One is whether the BZ and the manifold are compatible and the other is the condition of the Poincar\'{e}-Hopf theorem. These conditions depend on the concrete model. When these conditions are violated in some models such as the non-Hermitian torus model for some regions of the parameter space, what physics of the velocity field in the BZ and how do we connect the BZ and the manifold?  These questions are worth studying further.

It should be emphasized that the velocity field approach based on the connection between the velocity field and the Poincar\'{e}-Hopf theorem, is a new proposal. The velocity field approach is not a trivial application of the Poincar\'{e}-Hopf theorem because this approach involves the velocity field in the BZ and manifold of quantum models whether or not there exists a homeomorphism.

In addition, the scaling theory of $\mathbb{Z}_2$ topological invariants and the universal classes of topological phase transitions for high-order Dirac models was given.\cite{Chen2} A comparative study of the different physical properties of the topological invariants given by the velocity field flow and its scaling properties for higher dimensional systems is also worth while.

On the other hand, another important question is what physical meaning can be given to the topological invariants of the velocity field? In general, the Hall conductance depends on the current-current correlation function, based on linear response theory. \cite{Bernevig} Based on our existing work, the current depends on the velocity field in the BZ.  We believe that there should be some connection between conductance and the topological invariants of the velocity field.

\begin{acknowledgments}
The authors are grateful to Dr. Matthew J. LAKE for his improving the English presentation.
and also grateful to a referee for valuable comments and suggestions.
\end{acknowledgments}

\section{Appendixes}

\subsection{Poincar\'{e}-Hopf theorem}
We briefly review the Poincar\'{e}-Hopf theorem here. \cite{Lloyd}

\textbf{Poincar\'{e} theorem:} Let a vector field $\mathbf{X}$ be defined on a smooth closed two-dimensional Riemannian manifold $V$ and let it have a finite number of isolated singular points $x_1,\cdots, x_k$. Then
\begin{equation}\label{PT1}
\sum_{i}^{k}I(\mathbf{X},x_i)=\chi (V)
\end{equation}
where $I(\mathbf{X},x_i)$ is the index of the point $x_i$ with respect to $\mathbf{X}$ and $\chi$ is the Euler characteristic of $V$.

The index of the point $x_i$ with respect to $\mathbf{X}$ is defined in the following way.
Let a vector field $\mathbf{X}$ be defined on $\mathbb{R}^n$, and let $Q$ be a sphere of small radius surrounding a singular point $\mathbf{x}_0$ such that $\mathbf{X}|_Q\neq 0$.
The degree of the mapping  \cite{Lloyd}
\begin{equation}\label{DM1}
f:Q\rightarrow S^{n-1},    \quad f(\mathbf{x})=\frac{\mathbf{X}(\mathbf{x})}{\|\mathbf{X}(\mathbf{x})\|}
\end{equation}
is called the index, $I(\mathbf{X},\mathbf{x}_0)$, of the singular point $\mathbf{x}_0$ of the vector field $\mathbf{X}$, i.e.
\begin{equation}\label{Id1}
I(\mathbf{X},\mathbf{x}_0)=deg f_{\mathbf{x}_0}.
\end{equation}
If $\mathbf{x}_0$ is non-degenerate, then
\begin{equation}\label{Id2}
I(\mathbf{X},\mathbf{x}_0)=sign \det \left\|\frac{\partial \mathbf{X}^j}{\partial \mathbf{x}^i}\right\|.
\end{equation}
where $sign$ means the sign of the determinant.
This theorem was first established by H. Poincar\'{e} in 1881 and was generalized to the $n$-dimensional manifold by H. Hopf in 1926. \cite{Lloyd}

\textbf{Poincar\'{e}-Hopf theorem:}
Let $V$ be a compact orientable $n$-dimensional manifold, and let $\mathbf{X}$ be a continuous vector field with finitely many isolated singular points. Then the sum of the indices of those singular points is the Euler characteristic $\chi (V)$ of $V$. \cite{Eber,Lloyd}

\bibliography{apssamp}

\end{document}